\documentclass[letterpaper]{article} 
\usepackage[preprint]{aaai2027}  
\usepackage[hyphens]{url}  
\usepackage{graphicx} 
\usepackage{natbib}  
\usepackage{caption} 
\usepackage{algorithm}
\usepackage{algpseudocode}

\usepackage{newfloat}
\usepackage{listings}
\DeclareCaptionStyle{ruled}{labelfont=normalfont,labelsep=colon,strut=off} 
\floatstyle{ruled}
\newfloat{listing}{tb}{lst}{}
\floatname{listing}{Listing}

\usepackage{booktabs}

\usepackage{amsmath}
\usepackage{amssymb}
\usepackage{pgfplots}
\pgfplotsset{compat=1.18}
\usepackage{multirow}
\usepackage{tikz}
\usepackage[utf8]{inputenc}
\usetikzlibrary{shapes,arrows,positioning}
\usepackage{siunitx}

\title{\textsc{CoGate}: Confidence-Gated Co-Decoding for Secure Code Generation}

\title{\textsc{CoGate}: Confidence-Gated Co-Decoding for Secure Code Generation}
\author {
    Minghao Hu\textsuperscript{\rm 1} \thanks{This work is conducted during the internship at Thoughtworks.}
    Lannan Luo\textsuperscript{\rm 2},
    Allen Roush\textsuperscript{\rm 3}\corresponding
    Phillip Howard\textsuperscript{\rm 4}\corresponding
}
\affiliations {
    \textsuperscript{\rm 1}\textsuperscript{\rm 2}George Mason University, \quad
    \textsuperscript{\rm 3}\textsuperscript{\rm 4}Thoughtworks Inc.\\
    \{mhu20,lluo4\}@gmu.edu,\quad  \{allen.roush,phillip.howard\}@thoughtworks.com
}

\begin{document}

\maketitle

\begin{abstract}
Large language models are widely used for code generation, but they can also produce insecure programs due to patterns learned from their pretraining data. Decoding-time steering has become an important solution to this problem: a small expert model is combined with the target model at each step to generate more secure code, which is referred to as co-decoding.  However, the acceptance rule for existing co-decoding approaches does not consider the expert model's confidence. When the security expert is unconfident due to unseen patterns or out-of-distribution (OOD) contexts, its guidance can therefore be misleading. To address the challenge, we propose \textsc{CoGate}, a confidence-gated co-decoding approach that controls the expert's influence on the co-decoding process based on its confidence. We implement our approach and evaluate it across multiple LLM backends (CodeGen, DeepSeek-Coder, Qwen-Coder, StarCoder) on several code generation benchmarks (HumanEval, security suite, and CWEval). Our approach outperforms existing co-decoding methods (CoSec+) across multiple benchmarks, achieving up to a 12.6\% gain of Func-Sec@10 on CWEval.
\end{abstract}

\section{Introduction}
\label{sec:intro}

Large language models are widely utilized in code related tasks~\cite{shi2025between,shi2024code,shi2025longcodezip,park2024grammar,zhu2024hot,mundler2025type}, with popular LLM coding families such as GPT~\cite{achiam2023gpt}, Llama~\cite{touvron2023llama}, Qwen-Coder~\cite{hui2024qwen2}, DeepSeek-Coder~\cite{guo2024deepseek}, StarCoder~\cite{li2023starcoder}, and CodeGen~\cite{nijkamp2022codegen}. These code LLMs are pretrained on massive publicly collected code repositories, which may contain numerous insecure or low-quality code patterns that can be reflected in the model's generations. 
This poses a severe challenge considering the importance of generating secure code; programmers need to develop programs that are both functionally correct and safe from vulnerabilities~\cite{gong2025diffucoder}. Therefore, there is an urgent need to develop methods to guide LLMs to produce secure code without compromising its functionality.

Several approaches have been proposed for generating secure code, including Retrieval Augmented Generation (RAG)~\cite{shi2025rescue}, program repair~\cite{kong2025contrastrepair,yu2025patchagent} and sampling based methods~\cite{Cosec,CoSec+,kim-etal-2025-speak}. In the co-decoding paradigm, a small LLM trained on a security-related corpus infers the next token simultaneously with the base LLM. Acceptance rules are devised to choose whether the next token comes from the security LLM or the base LLM. 

Co-decoding approaches have two key advantages. First, they are training-free and operate at inference time, which allows them to be applied to the base model with the same tokenizer without retraining. Second, they do not require access to internal parameters of the base model, making them broadly applicable. Co-decoding methods such as CoSec~\cite{Cosec} and CoSec+~\cite{CoSec+}, have shown promising results in generating secure code by incorporating a security expert model to guide the base model during decoding. These methods make token-level decisions from a \emph{relative} comparison of the expert and the base model. At each step, the expert proposes a candidate token $\tilde{x}$, and a rule-based approach accepts it when $a \;<\; \min\!\left(1,\; B(\tilde{x})/S(\tilde{x})\right)$, where $B$ and $S$ are the base- and security-model probabilities of $\tilde{x}$ and $a$ is a fixed threshold. 

There are two key challenges associated with CoSec \& CoSec+: (1) Existing acceptance rules do not consider the token probability distribution of the security model. Therefore, a highly confident expert and an unconfident one are treated identically whenever they assign the same probability to $\tilde{x}$.
(2) For the acceptance rule, $\min(1,\,B_t(\tilde{x})/S_t(\tilde{x}))$, the acceptance probability \emph{increases} as the expert's sampled token probability decreases. Consequently, the less confidence the expert has in the token it draws, the more likely that token is accepted. 
This can commonly occur when the security expert model encounters OOD situations, causing its distribution becomes more uniform and long-tailed. In such cases, the selected token is fixed entirely by the base model and a uniform draw, carrying no security-relevant information at all. Therefore, co-decoding with an unconfident expert degenerates into simply noise injection. 


To address these challenges, we propose \textsc{CoGate}: a confidence-gated co-decoding method that measures the expert's confidence to decide whether it should be allowed to influence the generation. At each step of decoding, we compute a confidence signal $c(S_t)$ from the expert model's distribution and admit the expert's proposal only when it is above a given threshold $\tau$. When the expert model is unconfident ($c(S_t)<\tau$), the gate closes and sampling proceeds using only the base model's token probability distribution. When the expert is confident ($c(S_t)>\tau$), the gate is open and co-decoding proceeds using both models. 
We explore two confidence signals based on the expert model's token probability distribution: (1) the maximum probability among all the candidate tokens ($c(S_t)=\max S_t$), and (2) the distribution's overall dispersion as measured by Shannon entropy ($c(S_t)= 1-H(S_t) /\log |V|$), where $V$ is the vocabulary of candidate tokens. 

We evaluate \textsc{CoGate} on six models spanning four families (CodeGen, StarCoder, DeepSeek-Coder, and Qwen2.5-Coder) and compare two gating signals (max probability, normalized entropy) against the original base models, LoRA security fine-tuning, and the strongest prior co-decoding method (CoSec+). On HumanEval and Security Suite
benchmarks, \textsc{CoGate} outperforms CoSec+ in most settings. For StarCoder-7B, it achieves \textbf{54.5\%} on Pass@10 and \textbf{82.6\%} Security Ratio, which is an absolute improvement of \textbf{2.9\%} and \textbf{5.4\%} compared to CoSec+. On CWEval, the CWEs are unseen in previous training, and the multi-dimensional evaluation metric requires a single generation to be both functional and secure. For StarCoder-7B, \textsc{CoGate} achieves \textbf{35.2\%} on Func-Sec@10 for this task, a \textbf{2.4\%} absolute improvement over CoSec+. These out-of-distribution regimes are where the expert is most often unconfident, and thus where our gated co-decoding approach improves performance the most. 
Our contributions are summarized as follows:

\begin{itemize}
\item \textbf{A diagnosis of relative-confidence co-decoding.} We show both analytically and empirically (Section~\ref{sec:analysis}) that vanilla co-decoding (e.g., CoSec, CoSec+) conflates the expert's \emph{relative} preference with its \emph{qualification to steer}. In particular, its acceptance statistic is inverted in the expert's own confidence, causing an under-confident expert to be accepted more readily. 

\item \textbf{\textsc{CoGate}.} We propose \textsc{CoGate}, a confidence-modulated co-decoding scheme. It measures the expert model's confidence and uses it to modulate the authority of the expert's intervention, separating the relative preference from the absolute confidence.


\item \textbf{Intervention Utility Analysis.} We analyze the gain when expert steering is applied, quantifying how the expert's confidence modulates the intervention utility. We find that there exists a range of thresholds where the expert's confidence reliably predicts whether its intervention will be beneficial, guiding when the gate should open.

\item \textbf{Comprehensive Evaluation.} We conduct extensive experiments across six models on three benchmarks. The evaluation results show that \textsc{CoGate} outperforms CoSec+ and the LoRA fine-tuned base model in most settings, proving the effectiveness of our approach.
\end{itemize}
\section{Related Work}
\label{sec:related-work}

\subsection{Secure Code Generation}

Security of software program is critival in software development~\cite{ma2026llm,duan2026meta,wang2026model,hu2026zero,hu2025flow}, which incurs the need for secure code generation~\cite{nie2025decoding,nazzal2024promsec,xu2024safedecoding,nie2025decoding,light2025sfs}. These findings motivated a line of \emph{security hardening} methods that steer generation toward secure code. SVEN~\cite{he2023large} learns a pair of prefix vectors that push generation toward secure or vulnerable outputs without modifying model weights. SafeCoder~\cite{he2024instruction} folds security-centric fine-tuning into instruction tuning to jointly optimize security and utility. 
CoSec~\cite{Cosec} introduced co-decoding, in which a small security model reasons alongside the target at inference and, through a confidence-based acceptance rule, nudges it toward secure tokens without touching its parameters. CoSec+~\cite{CoSec+} strengthens this pipeline with knowledge distillation and improved security training; it is the method we build on and compare against. Existing works, nonetheless, do not evaluate program security and functional correctness at the same time. They evaluate these two aspects separately, ignoring whether a generated program is both secure and correct. 

A parallel and equally important thread concerns \emph{evaluation}. Most security benchmarks score only whether generated code is free of a target weakness, ignoring whether it still works. CodeGuard+~\cite{fu2024constrained} showed that this is misleading: measuring security in isolation can substantially overestimate a defense's value, and a leading hardening method (prefix tuning) improves security while sacrificing functional correctness. CWEval~\cite{peng2025cweval} pushes this further with an outcome-driven, multilingual benchmark whose executable oracles verify functionality and security simultaneously. 

\subsection{Decoding-Time Steering}

Adjusting a frozen model's output at inference by combining it with an auxiliary signal is a well-established paradigm for controllable generation~\cite{wang2025secdecoding}. DExperts~\cite{liu2021dexperts} reweights a base model's logits using an expert and an anti-expert. FUDGE~\cite{yang2021fudge} multiplies the base distribution by an attribute predictor operating on partial sequences. Critic-guided decoding~\cite{kim2023critic} uses a reward-trained critic to reweight the output distribution. CoSec and CoSec+ belong to this family, and its accept-or-resample mechanism is closely related to speculative decoding~\cite{leviathan2023fast}, in which a small draft model proposes tokens that a larger model verifies. What these methods share is that their per-step control signal is \emph{relative}: a ratio or difference between the expert and the base. Such a signal captures how strongly the expert prefers a token over the base, but is by construction insensitive to whether the expert is a confident, informative guide at that step. Our work targets exactly this blind spot in the co-decoding setting.

\section{Motivation}
\label{sec:motivation}

\subsection{Preliminaries: Co-Decoding for Security}
\label{sec:prelim}
 
Let $\mathcal{V}$ be the vocabulary and $V=|\mathcal{V}|$. At decoding step $t$ with context $x_{<t}=(x_1,\dots,x_{t-1})$, the frozen target (base) model induces a next-token distribution $B_t(\cdot)=B(\cdot\mid x_{<t})$ over $\mathcal{V}$, and a small security expert induces $S_t(\cdot)=S(\cdot\mid x_{<t})$. Following CoSec+, the expert is a compact model (functional-correctness aligned by distillation, then security-trained) that places higher probability on secure tokens.
Co-decoding proceeds one token at a time. The expert proposes a candidate $\tilde{x}\sim S_t$, and the candidate is accepted as the output iff $a \;<\; \min\!(1,\ B_t(\tilde{x})/S_t(\tilde{x}))$, where $a\in[0,1)$ is a fixed acceptance threshold (default $a=0.3$). If the test fails, the step resamples from the base, $x_t\sim B_t$. CoSec+ admits the expert's secure token unless the base assigns it disproportionately low probability, i.e., unless the base ``strongly disagrees.'' The security signal enters only through $S_t$.
 
\subsection{The Failure Mode: Confidence Miscalibration}
\label{sec:analysis}
 
The acceptance criteria in CoSec+ lacks any measure of the expert model's confidence (e.g., absolute confidence):
\begin{equation}
c_\text{max}(S_t) \;:=\; \max_{x\in\mathcal{V}} S_t(x),
\end{equation}
 This leads to two undesirable properties: confidence inflation and security-signal collapse.
 
\medskip
\noindent\textbf{Proposition 1 (Confidence inflation).} \emph{Fix a candidate $\tilde{x}$ with $B_t(\tilde{x})>0$. The acceptance statistic $\rho:=\min\!\big(1,\,B_t(\tilde{x})/S_t(\tilde{x})\big)$ is nonincreasing in $S_t(\tilde{x})$. Hence the lower the expert's probability on the very token it proposes, the more readily that token is accepted.}
 
\smallskip
\noindent\emph{Proof.} For fixed $B_t(\tilde{x})>0$, the map $s\mapsto B_t(\tilde{x})/s$ is decreasing on $(0,\infty)$, and $u\mapsto\min(1,u)$ is nondecreasing, so $s\mapsto\min\!\big(1,B_t(\tilde{x})/s\big)$ is nonincreasing. The event $\{a<\rho\}$ is therefore weakly easier to satisfy as $S_t(\tilde{x})$ shrinks. $\square$
 
This is problematic because reliability should raise the bar for acting on the expert, not lower it. An unconfident expert assigns small probability to whichever token it samples, which by Proposition~1 makes acceptance \emph{more} likely precisely when the proposal is least trustworthy. The next result shows what is being accepted in the limit.
 
\noindent\textbf{Proposition 2 (Security-signal collapse).} \emph{Suppose the expert is unconfident at step $t$, i.e.\ $S_t\sim U$, the near-uniform distribution on $\mathcal{V}$, and $\tilde{x}\sim S_t$. Then for $a<1$ the accept event is $\{B_t(\tilde{x})>a/V\}$, and both the accept decision and the emitted token depend only on $B_t$ and a near-uniform draw: never on any security-relevant property of the context. Marginalizing over the draw, the probability of emitting the expert's context-independent token is}
\begin{equation}
\label{eq:pacc}
\begin{split}
p_{\mathrm{acc}} 
&\;=\; \frac{1}{V}\,\big|\{x\in\mathcal{V}: B_t(x)>a/V\}\big|.
\end{split}
\end{equation}
 
\smallskip
\noindent\emph{Proof.} If $S_t \sim U$ then $\tilde{x}$ is near-uniform on $\mathcal{V}$ and $S_t(\tilde{x})=1/V$ for every outcome, so $\rho=\min(1,V\,B_t(\tilde{x}))$. For $a<1$, $\rho>a \iff V\,B_t(\tilde{x})>a \iff B_t(\tilde{x})>a/V$. $S_t$, which is the only security-trained component, is constant. So neither the indicator $\mathbf{1}[\rho>a]$ nor the accepted token $\tilde{x}$ carries information about the context beyond what $B_t$ supplies. Equation~\eqref{eq:pacc} follows by taking the expectation of $\mathbf{1}[B_t(\tilde{x})>a/V]$ under the near-uniform draw. $\square$
 
\noindent\textbf{Remark (Co-uncertainty).} By \eqref{eq:pacc}, the rate at which the mechanism injects the expert's noise is governed by the \emph{base} model's concentration. When $B_t$ is sharply peaked, few tokens clear the small threshold $a/V$, so $p_{\mathrm{acc}}$ is small and generation harmlessly defers to $B_t$. When $B_t$ is itself diffuse, $p_{\mathrm{acc}}$ grows and the emitted token is effectively random. The damaging regime is thus \emph{joint} uncertainty: both $S_t$ and $B_t$ are flat. This is exactly the profile of ambiguous, long-tail, and out-of-distribution decision points, and these are disproportionately the security-decisive ones. 
Standard, in-distribution CWEs on which the expert is confident do not trigger the failure, which is why disjoint security/correctness benchmarks that emphasize such cases leave it hidden.
 
\noindent\textbf{Interpretation.} Proposition~2 is stated in the idealized limit $S_t \sim U$. In practice the expert merely flattens rather than reaching uniformity. Proposition~1 shows the effect is \emph{monotone}, so it is already active before the exact limit: as $c(S_t)$ falls, the expert's probability on its own proposal falls with it, the ratio inflates, and acceptance decouples from security. 
The corrective, then, is to condition the expert's intervention on $c(S_t)$ , which we subsequently propose in the next section.

\begin{figure}[t]
    \centering
    \includegraphics[width=\linewidth]{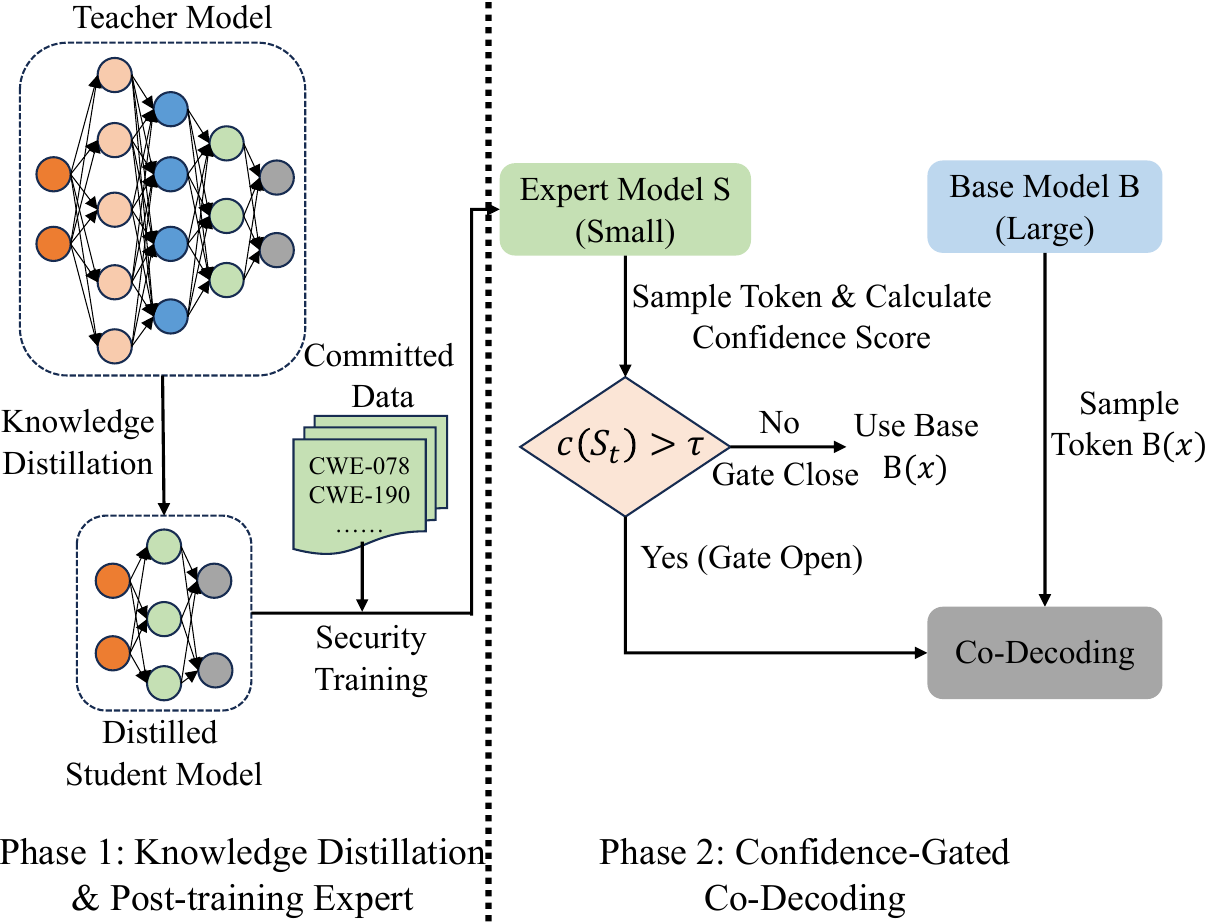}
    \caption{Overview of the \textsc{CoGate}.}
    \label{fig:cogate_overview}
\end{figure}

\section{Model Design}
\label{sec:model-design}

\subsection{Overview}
To address the failure mode identified in Section~\ref{sec:analysis}, we propose \textsc{CoGate}, a confidence-gate mechanism of co-decoding that explores multiple metrics to consider whether the expert is a reliable guide upon the decoding step. 
Figure~\ref{fig:cogate_overview} presents the overview of \textsc{CoGate}, which consists of two phases: (1) Knowledge distillation \& post-training expert, and (2) Confidence-Gated Co-Decoding. In Phase 1, we follow the same procedure in CoSec+, including distilling knowledge from the base model into the expert and post-training the expert to improve its security signal, so that it can later provide reliable guidance during co-decoding. In phase 2, the expert proposes tokens as usual, but the confidence gate decides whether the expert is allowed to influence the output at each step, ensuring that only sufficiently confident expert interventions are considered. 
 
\subsection{Confidence-Gated Co-Decoding}
\label{sec:gate}
 
\textsc{CoGate} adds an \emph{confidence gate} that decides, before the acceptance criteria is computed, whether the expert should be used at all for co-decoding at the current generation step. Using the maximum probability gate as an example, we define the gate as:
\begin{equation}
\label{eq:gate}
g_t \;=\; \mathbf{1}\!\left[\,c(S_t) > \tau\,\right],
\qquad c(S_t):=\max_{x\in\mathcal{V}}S_t(x),
\end{equation}
with a single hyperparameter $\tau\in[0,1]$. $c(S_t)$ measures the expert's confidence at step $t$. The gated acceptance rule is
\begin{equation}
\label{eq:gated-accept}
\text{accept }\tilde{x}
\iff
\underbrace{g_t=1}_{\text{absolute reliability}}
\ \wedge\
\underbrace{a<\min\!\big(1,\,B_t(\tilde{x})/S_t(\tilde{x})\big)}_{\text{relative preference}} .
\end{equation}
If $\tilde{x}$ is not accepted, whether because the gate is closed or because the ratio test fails, the step resamples from the base, $x_t\sim B_t$. The two conjuncts in \eqref{eq:gated-accept} are the two signals we aim to separate: the ratio measures how much the expert prefers $\tilde{x}$ over the base, while the gate independently certifies that the expert is confident enough for that preference to be meaningful. The pathological scenario described in Section~\ref{sec:analysis} (flat $S_t$, hence tiny $S_t(\tilde{x})$ and an inflated ratio) now has a small $c(S_t)$, so $g_t=0$ and generation defers cleanly to $B_t$. When the expert is confident and the base concurs, $g_t=1$ and co-decoding proceeds exactly as before.
 
\begin{algorithm}[t]
\caption{Confidence-Gated Co-Decoding (\textsc{CoGate})}
\label{alg:cogate}
\begin{algorithmic}[1]
\Require prompt $x_{1:n_0}$; Base $B$; Expert $S$; acceptance threshold $a$; gate threshold $\tau$; confidence functional $c(S_t)$ ; max length $L$
\While{$t<L$ \textbf{and} $x_t \neq \textsc{eos}$}
  \State $S_t \gets S(\cdot\mid x_{1:t});\quad B_t \gets B(\cdot\mid x_{1:t})$
  \State $c(S_t) \gets S$ \Comment{calculate expert confidence}
  \If{$c(S_t) > \tau$} \Comment{gate open}
    \State $\tilde{x} \sim S_t$ \Comment{expert proposes}
    \If{$a < \min\!\big(1,\,B_t(\tilde{x})/S_t(\tilde{x})\big)$}
      \State $x_{t+1} \gets \tilde{x}$ \Comment{accept secure token}
    \Else
      \State $x_{t+1} \sim B_t$ \Comment{resample from base}
    \EndIf
  \Else \Comment{gate closed: expert abstains}
    \State $x_{t+1} \sim B_t$ \Comment{defer to base}
  \EndIf
  \State $t \gets t+1$
\EndWhile
\State \Return $x_{1:t}$
\end{algorithmic}
\end{algorithm}
 
Algorithm~\ref{alg:cogate} details the full procedure. Two properties of this approach are worth noting. First, \textsc{CoGate} \emph{strictly generalizes} CoSec+: since $c(S_t)> 1/V$ always, setting $\tau\le 1/V$ leaves the gate permanently open and recovers CoSec+ exactly, while $\tau\!\to\!1$ closes it on all but fully deterministic expert steps, approaching pure base-model decoding. The threshold $\tau$ thus interpolates between the original method and no steering. Second, the gate adds only the cost of computing $c(S_t)$ from logits already produced by the expert model, so per-step overhead is negligible and the co-decoding inference process is otherwise unchanged.
 
\subsection{Confidence Gates}
\label{sec:family}
 
The maximum probability $c_\text{max}(S_t)$ is one way to certify that the expert is informative, but not the only one. We frame the gate through a \emph{confidence functional} $c(S_t)$ and open it when $c(S_t)\ge\tau$ (for uncertainty measures, the inequality reverses). This exposes a design space that we study empirically, with each confidence functional evaluating a different aspect of the expert's distribution.
\begin{itemize}
  \item \textbf{Max-probability}: $c_{\text{max}}(S_t)=\max_x S_t(x)$. Gates on the probability of the most likely token.
  \item \textbf{Normalized Entropy}: $c_{\text{ent}}(S_t)=1-H(S_t)/\log |V|$. Characterizes the sharpness of the entire distribution rather than a single coordinate, and is the natural information-theoretic measure of the flatness that drives Proposition~2.
\end{itemize}
All variants are computable in a single pass over the expert's logits and leave the rest of the pipeline untouched. Section~\ref{sec:experiments} reports which signal best separates informative from uninformative expert steps.

\subsection{The False-Accept/False-Reject Trade-off}
\label{sec:tradeoff}
 
The gate turns the expert into a \emph{selective} guide that abstains when unsure, with $\tau$ mediating its behavior. To characterize what $\tau$ controls, consider an idealized oracle that labels a step's expert proposal $\tilde{x}$ \emph{reliable} (event $R_t$) when it is both secure and consistent with a functionally correct continuation. Two errors then arise at the token level:
\begin{itemize}
  \item \textbf{False accept} (FA): The rule emits $\tilde{x}$ although $\neg R_t$ acts on an unreliable, typically noisy, proposal. FAs chiefly damage functional correctness (the emitted token need not compile or fit the context) and can also introduce vulnerabilities.
  \item \textbf{False reject} (FR): A genuinely reliable secure proposal ($R_t$) is suppressed, either because the gate is closed or because the ratio test fails. FRs chiefly weaken security, since a correct secure suggestion is discarded in favor of the base model's token.
\end{itemize}
CoSec+ has no gate ($\tau\le 1/V$) and therefore incurs the full FA rate implied by Proposition~2. Raising $\tau$ trades one error for the other, monotonically.

 
\section{Evaluation}
\label{sec:experiments}

\begin{table*}[ht]
\centering
\setlength{\tabcolsep}{3mm}
\begin{tabular}{
    c|l|
    S[table-format=2.1]
    S[table-format=2.1]
    S[table-format=2.1]
    S[table-format=2.1]
    S[table-format=2.1]
}
\hline
\multirow{2}{*}{Model}
& \multirow{2}{*}{Type}
& \multicolumn{2}{c}{HumanEval}
& \multicolumn{1}{c}{Security Suite}
& \multicolumn{2}{c}{CWEval} \\
\cline{3-7}
&
& \multicolumn{1}{c}{Pass@1}
& \multicolumn{1}{c}{Pass@10}
& \multicolumn{1}{c}{Ratio}
& \multicolumn{1}{c}{Func-Sec@1}
& \multicolumn{1}{c}{Func-Sec@10} \\
\hline
\hline

\multirow{5}{*}{\shortstack{CodeGen\\-2.7B}}
& Original  & 8.3 & 44.8 & 52.7 & 7.9 & 27.5 \\
& +LoRA     & 8.1 & 47.2 & 53.8 & 9.2 & 29.4 \\
& CoSec+   & 9.4 & 48.1 & 56.2 & 9.8 & 30.8 \\
& \textsc{CoGate} (Max-Prob)  & {\bfseries 10.2} & 50.4
            & 66.2 & {\bfseries 13.7} & 32.6 \\
& \textsc{CoGate} (Entropy)   & 9.2 & {\bfseries 50.9}
            & {\bfseries 66.5} & 10.9 & {\bfseries 33.1} \\
\hline

\multirow{5}{*}{\shortstack{CodeGen\\-6.1B}}
& Original  & 17.0 & 46.3 & 61.3 & 10.5 & 34.6 \\
& +LoRA     & 16.8 & 44.0 & 63.4 & 12.8 & 37.7 \\
& CoSec+   & 14.0 & 49.8 & 57.2 & 11.2 & 36.8 \\
& \textsc{CoGate} (Max-Prob)  & 13.9 & 52.2
            & 62.2 & {\bfseries 14.4} & 44.0 \\
& \textsc{CoGate} (Entropy)   & {\bfseries 17.7} & {\bfseries 52.7}
            & {\bfseries 64.5} & 12.6 & {\bfseries 44.4} \\
\hline

\multirow{5}{*}{\shortstack{StarCoder\\-1B}}
& Original  & 11.6 & 23.5 & 40.9 & 8.1 & 27.9 \\
& +LoRA     & 12.4 & 21.8 & 54.9 & 9.3 & 29.7 \\
& CoSec+   & 16.7 & 35.7 & 56.4 & 10.0 & 31.0 \\
& \textsc{CoGate} (Max-Prob)  & {\bfseries 18.5} & 41.0
            & {\bfseries 58.3} & 10.9 & 29.6 \\
& \textsc{CoGate} (Entropy)   & 16.4 & {\bfseries 45.4}
            & 37.7 & {\bfseries 11.1} & {\bfseries 33.3} \\
\hline

\multirow{5}{*}{\shortstack{StarCoder\\-7B}}
& Original  & {\bfseries 19.5} & 48.0 & 70.8 & 10.9 & 29.6 \\
& +LoRA     & 18.2 & 50.7 & 65.9 & 12.2 & 31.7 \\
& CoSec+   & 16.5 & 51.6 & 77.2 & 12.7 & 32.8 \\
& \textsc{CoGate} (Max-Prob)  & 17.4 & 54.1
            & 79.3 & 11.9 & 32.8 \\
& \textsc{CoGate} (Entropy)   & 15.2 & {\bfseries 54.5}
            & {\bfseries 82.6} & {\bfseries 13.1}
            & {\bfseries 35.2} \\
\hline

\multirow{5}{*}{\shortstack{DeepSeek\\-Coder-6.7B}}
& Original  & 20.2 & 69.1 & 72.1 & 9.2 & 30.2 \\
& +LoRA     & 21.5 & 65.8 & 67.3 & 10.5 & 32.2 \\
& CoSec+   & 22.0 & 67.7 & 71.5 & 11.0 & 33.4 \\
& \textsc{CoGate} (Max-Prob)  & {\bfseries 23.7} & {\bfseries 75.0}
            & {\bfseries 80.5} & 12.2 & 35.7 \\
& \textsc{CoGate} (Entropy)   & 23.4 & 72.4
            & 76.6 & {\bfseries 12.4} & {\bfseries 36.1} \\
\hline

\multirow{5}{*}{\shortstack{Qwen2.5\\-Coder-14B}}
& Original  & 70.2 & 86.6 & 69.6 & 13.6 & 41.0 \\
& +LoRA     & 69.9 & 82.2 & 68.2 & 14.9 & 43.0 \\
& CoSec+   & 68.2 & 85.0 & 68.0 & 16.4 & 44.2 \\
& \textsc{CoGate} (Max-Prob)  & {\bfseries 73.2} & {\bfseries 87.4}
            & 69.7 & 16.6 & 49.4 \\
& \textsc{CoGate} (Entropy)   & 70.9 & 86.8
            & {\bfseries 70.5} & {\bfseries 19.8}
            & {\bfseries 56.8} \\
\hline

\end{tabular}
\caption{Evaluation results comparing different models and methods on HumanEval, Security Suite, and CWEval benchmarks, showing functional correctness and security performance.}
\label{tab:result}
\end{table*}

We conduct comprehensive experiments to address the following research questions (\textbf{RQ}s):
\begin{itemize}
  \item \textbf{RQ1 (Effectiveness).} How does \textsc{CoGate} perform against existing baselines? 
  \item \textbf{RQ2 (Intervention Utility).} What is the utility gain of \textsc{CoGate} in suppressing unconfident interventions and how sensitive is it to the confidence threshold $\tau$? 
  \item \textbf{RQ3 (Gate Choice).} Which confidence signal most reliably identifies harmful expert interventions?
  \item \textbf{RQ4 (Temperature).} How do different temperatures affect the performance of \textsc{CoGate}?
\end{itemize}

\subsection{RQ1: Effectiveness}
\label{sec:rq1}

\noindent \textbf{LLM Backends.}
We evaluate six target models across four families: CodeGen (2.7B, 6.1B), StarCoderBase (1B, 7B), DeepSeek-Coder (6.7B), and Qwen2.5-Coder (14B). For each model family, we choose the smallest model (350M for CodeGen, 1B for StarCoderBase, 1.3B for DeepSeek-Coder, and 1.5B for Qwen2.5-Coder) as the security expert. For the security expert model, we follow the same knowledge distillation and fine-tuning procedure proposed by CoSec+. One exception is for StarCoder-1B, which serves both as the security expert model and target model; therefore, we skip its knowledge distillation step. We evaluate each of the two types of inference-time confidence gate mechanisms described in Section~\ref{sec:family}. 
 
\noindent \textbf{Baselines.}
We compare \textsc{CoGate}, including  two gating signals against (1) \textbf{\emph{Original}}, the unmodified target model; (2) \textbf{\emph{LoRA}}, security fine-tuning applied directly to the target model rather than the expert model. We follow the same procedure in CoSec+ to fine-tune the target model; (3) \textbf{\emph{CoSec+}}~\cite{CoSec+}, the strongest prior method and our primary point of comparison. 
 
\noindent \textbf{Benchmarks and metrics.}
We use three evaluation benchmarks. (1) The \textbf{\emph{security suite}} used in SVEN~\cite{he2023large} that measures the \emph{security ratio} (fraction of compilable, de-duplicated generations judged secure by CodeQL) over the CWE types seen in the expert's training corpus; this is our in-distribution security probe. (2) \textbf{\emph{HumanEval}}~\cite{chen2021evaluating} measures functional correctness via Pass@$k$ ($k\in\{1,10\}$). (3) \textbf{\emph{CWEval}}~\cite{peng2025cweval} is an outcome-driven, multilingual (C, C++, Go, JavaScript, Python) benchmark whose executable oracles score functionality and security \emph{simultaneously}. We use the metric Func-Sec@$k$ ($k\in\{1,10\}$), which is the fraction of generations that are both functionally correct and free of the target weakness. Because CWEval spans languages and scenarios sparsely represented in the compact secure-coding corpus, it is our primary OOD setting. 
 
\noindent \textbf{Implementation.}
For each pair of LLMs, we use two gating mechanisms (max-probability, entropy) to decide whether the expert should intervene. We set the gate threshold $\tau$ per family on a held-out development split by maximizing CWEval Func-Sec@10, then freeze it for all reported results. The sensitivity analysis for $\tau$ is included in Section~\ref{sec:rq2}. Other inference hyperparameters include: 25 samples per prompt, up to 256 new tokens, top-$p=0.95$, sampling temperature $0.6$, acceptance threshold $a=0.3$. All experiments were run on Nvidia H100 GPUs.

\noindent \textbf{Overview of Results.}
Table~\ref{tab:result} reports the results of two confidence gating mechanisms (max-probability and normalized entropy) along with baseline methods (Original, LoRA, CoSec+) across three benchmarks for multiple models. For the HumanEval and security suite, \textsc{CoGate} outperforms CoSec+ in most settings.

On CWEval, the CWEs are unseen in the post-training. Therefore, the gain achieved by \textsc{CoGate} in Func-Sec@k is attributable to the mechanism of the confidence-gated expert: when the expert is unconfident in unseen CWEs, it has a relatively low $c(S_t)$ for OOD inputs. This causes the gate to intervene, preventing the expert from providing harmful guidance. Hence, \textsc{CoGate} helps rule out the insecure and incorrect generations that would degrade performance.

\subsection{RQ2: Intervention Utility \& Sensitivity Analysis}
\label{sec:rq2}

To answer \textbf{RQ2}, we conduct a token-level counterfactual intervention experiment. This experiment directly validates our theoretical analysis (Proposition 2) by isolating the impact of individual expert tokens on the security and correctness of final program.

\noindent \textbf{Experimental Setup.} 
To avoid prefix confounding, where different sampled tokens lead to entirely different subsequent contexts, we collect a fixed set of generation prefixes $h_t$ from base-model trajectories. We use StarCoder-7B as the base model with the distilled and fine-tuned StarCoder-1B as the security expert model, evaluating both of our two confidence signals. For each step where the confidence of security expert $c(S_t)$ passes over the predetermined threshold $\tau$, we record the expert's absolute confidence $c(S_t)$ and fork the generation into three parallel branches: (1) \emph{\textsc{CoGate} branch:} the decoding follows the CoSec+ given the expert is confident, (2) \emph{Unconfident CoSec+ branch:} the decoding follows the CoSec+ given the expert is unconfident, and (3) \emph{Base branch:} we force the decoding to emit the base model's preferred token regardless of the expert's proposal.

From these divergent tokens, we resume generation using identical decoding policies and paired random seeds. To reduce the variance of single continuations, we sample $K=25$ paired continuations per branch and evaluate their security and functional correctness on CWEval. We then compute the average probability of passing the security and functional correctness tests for three branches, denoted as $\mathcal{P}_1$, $\mathcal{P}_2$, and $\mathcal{P}_3$. For example, $\mathcal{P}_1$ is the probability of passing the security and functional correctness tests for the \textsc{CoGate} branch: 

\begin{align}
    \mathcal{P}_1
    &= \frac{1}{K}\sum\nolimits_{k=1}^{K}
    \mathbf{1}\!\left[
        \text{Func-Sec}\!\left(
        x^{(k)}_{\textsc{CoGate}}
        \right)
    \right] \\
    \mathcal{P}_2
    &= \frac{1}{K}\sum\nolimits_{k=1}^{K}
    \mathbf{1}\!\left[
        \text{Func-Sec}\!\left(
        x^{(k)}_{\text{Unconfident CoSec+}}
        \right)
    \right] \\
    \mathcal{P}_3
    &= \frac{1}{K}\sum\nolimits_{k=1}^{K}
    \mathbf{1}\!\left[
        \text{Func-Sec}\!\left(
        x^{(k)}_{\text{Base}}
        \right)
    \right]
\end{align}

We define the \textit{token-level intervention utility} $U$ of \emph{\textsc{CoGate} branch} and \emph{Unconfident CoSec+ branch} against the \emph{Base branch} as the difference in pass rates:

\begin{align}
    U_{\textsc{CoGate}\text{-Base}}
    &= \mathcal{P}_1 - \mathcal{P}_3, \\
    U_{\text{Unconfident CoSec+}\text{-Base}}
    &= \mathcal{P}_2 - \mathcal{P}_3.
\end{align}

Thus, $U< 0$ indicates that the intervention is harmful compared to the base model, $U> 0$ means it is helpful compared to the base model, and $U_t \approx 0$ implies neutrality.

\begin{figure}[ht]
    \centering
    \includegraphics[width=\linewidth]{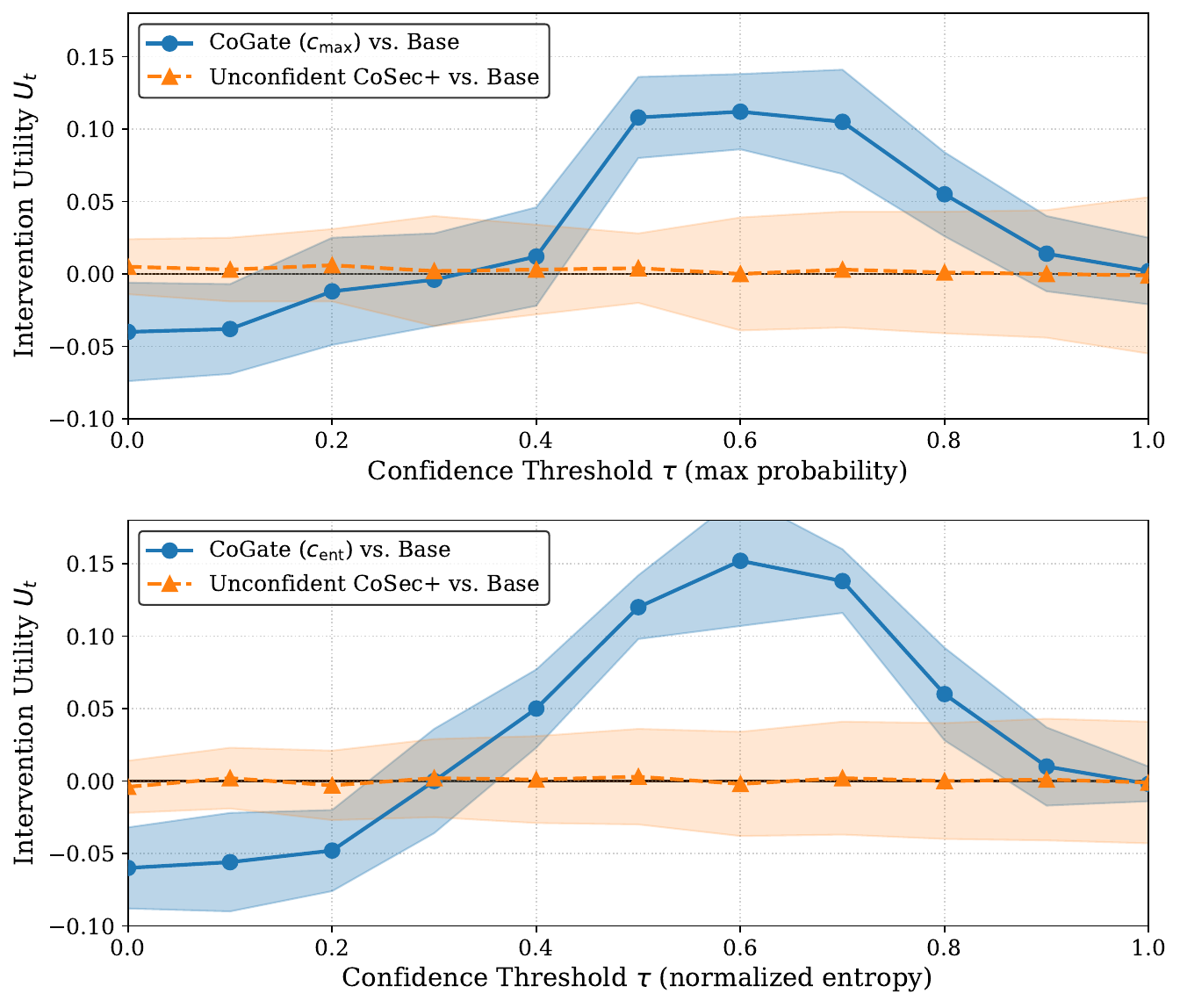}
    \caption{Token-level intervention utility $\mathbb{E}[U]$ for two types of confidence signals (max-probability, normalized entropy). 
    The shaded area represents the 95\% confidence interval.}
    \label{fig:intervention_utility}
\end{figure}

\noindent \textbf{Results.}
Figure~\ref{fig:intervention_utility} shows the expected intervention utility $\mathbb{E}[U_t]$ across multiple confidence levels $\tau$ using the two signals. This analysis yields the following findings: (1) For StarCoder-7B, setting max-probability threshold around 0.5 to 0.7 yields a satisfactory performance gain and intervention utility. Similar trends are observed for normalized entropy confidence signal. (2) Setting the threshold too low (e.g., 0.1) leads to a negative intervention utility, indicating that the expert's guidance is more likely to be harmful than helpful. (3) With a threshold higher than 0.9, the intervention utility reaches zero.  This means that the expert is nearly completely silenced and \textsc{CoGate} degrades to the base model. The intervention utility of unconfident CoSec+ becomes unstable as the confidence threshold increases. This shows that allowing an unconfident expert to steer decoding can lead to inconsistent outcomes, validating the theoretical analysis in Proposition 2.

\noindent \textbf{Sensitivity Analysis.}
We analyze the sensitivity of the model's performance to different values of this threshold. From Figure~\ref{fig:intervention_utility} we can observe the following: (1) The composite metric traces an inverted-U shape. It rises from CoSec+ value at $\tau = 0$, surpassing the base scenario at an intermediate threshold (0.35 for max probability and 0.32 for normalized entropy). (2) The performance reaches its peak around 0.6 and then declines as $\tau$ increases further, indicating that there is an optimal range (around 0.5 to 0.7) for the gate threshold that balances the expert's influence. (3) The shape of the sensitivity curve is consistent with Proposition 2 in Section~\ref{sec:analysis}: there are harmful interventions at low expert confidence, which can be removed while raising the threshold $\tau$. The security degrades gracefully while functional correctness improves so that the composite Func-Sec utility peaks before the expert is fully silenced near $\tau=1$.

\subsection{RQ3: Confidence Signal Choice}
\label{sec:rq3}

In this section, we investigate the impact of the two different confidence signals: max-probability and normalized entropy. We compare the two signals in a series of metrics.


\noindent \textbf{Scenario-specific Behaviors.}
 The two confidence signals behave the same way for two extreme scenarios: uniform distribution ($c_\text{max} \rightarrow 1/|V|$, $c_\text{ent} \rightarrow 0$) or one-hot distribution ($c_\text{max} \rightarrow 1$, $c_\text{ent} \rightarrow 1$).
 For intermediate scenarios, the two confidence signals can behave differently. For instance, when the distribution has two dominant peaks (0.45, 0.4, 0.05, 0.03...), $c_\text{max}=0.45$ while $c_\text{ent}$ is relatively high since it's far from uniform. Therefore, max probability tends to close the gate while normalized entropy may keep it open. On the other hand, if the distribution has one dominant peak and several smaller ones (0.8, 0.1, 0.05, 0.05...), $c_\text{max}=0.8$ while $c_\text{ent}$ is lower. As a result, max probability tends to trust this expert while the normalized entropy gate tends to be cautious. 

\noindent \textbf{Recommendation.}
Max-probability is parameter-light, transparent and easy to implement across vocabularies and LLM backends. Entropy is preferable where the two gates may diverge in jointly-scored regimes and unseen CWEs where the OOD and long tail distributions matter. This conclusion is consistent with the results in Table~\ref{tab:result} that using normalized entropy outperforms max-probability in CWEval unseen CWEs for most cases. Therefore, we recommend using max-probability for in-distribution scenarios and normalized entropy for OOD scenarios.

\subsection{RQ4: Temperature Study}

\begin{table}[ht]
\centering
\small
\setlength{\tabcolsep}{1mm}
\begin{tabular}{
    c|l|
    S[table-format=2.1, detect-weight=true]
    S[table-format=2.1, detect-weight=true]
    S[table-format=2.1, detect-weight=true]
    S[table-format=2.1, detect-weight=true]
}
\hline
\multirow{2}{*}{Model}
& \multirow{2}{*}{Type}
& \multicolumn{4}{c}{CWEval Func-Sec@10 (\%)} \\
\cline{3-6}
&
& \multicolumn{1}{c}{T=0.4}
& \multicolumn{1}{c}{T=0.6}
& \multicolumn{1}{c}{T=0.8}
& \multicolumn{1}{c}{T=1.0} \\
\hline
\hline

\multirow{5}{*}{\shortstack{CodeGen\\-2.7B}}
& Original   & 26.2 & 27.5 & 27.9 & 25.8 \\
& +LoRA      & 28.1 & 29.4 & 29.0 & 26.9 \\
& CoSec+     & 30.1 & 30.8 & 29.3 & 26.4 \\
& \textsc{CoGate} (Max-Prob)  & \textbf{31.6} & 32.6 & 28.5 & 29.7 \\
& \textsc{CoGate} (Entropy)    & 31.2 & \textbf{33.1}
             & \textbf{34.8} & \textbf{33.9} \\
\hline

\multirow{5}{*}{\shortstack{CodeGen\\-6.1B}}
& Original   & 33.2 & 34.6 & 35.1 & 24.9 \\
& +LoRA      & 36.1 & 37.7 & 37.1 & 28.8 \\
& CoSec+     & \textbf{38.7} & 36.8 & 34.9 & 26.8 \\
& \textsc{CoGate} (Max-Prob)   & 37.8 & 44.0 & 39.1 & \textbf{38.6} \\
& \textsc{CoGate} (Entropy)    & 38.0 & \textbf{44.4}
             & \textbf{42.0} & 37.8 \\
\hline

\multirow{5}{*}{\shortstack{StarCoder\\-1B}}
& Original   & 26.3 & 27.9 & 27.2 & 24.8 \\
& +LoRA      & 28.0 & 29.7 & 29.1 & 26.2 \\
& CoSec+     & 30.4 & 31.0 & 29.5 & 26.8 \\
& \textsc{CoGate} (Max-Prob)   & 29.3 & 29.6 & 28.7 & 26.1 \\
& \textsc{CoGate} (Entropy)    & \textbf{31.5} & \textbf{33.3}
             & \textbf{32.8} & \textbf{32.2} \\
\hline

\multirow{5}{*}{\shortstack{StarCoder\\-7B}}
& Original   & 28.2 & 29.6 & 30.1 & 28.0 \\
& +LoRA      & 30.3 & 31.7 & 31.2 & 28.9 \\
& CoSec+     & 32.1 & 32.8 & 31.0 & 27.6 \\
& \textsc{CoGate} (Max-Prob)   & 32.4 & 32.8 & 31.9 & 32.1 \\
& \textsc{CoGate} (Entropy)    & \textbf{33.0} & \textbf{35.2}
             & \textbf{36.9} & \textbf{35.7} \\
\hline

\multirow{5}{*}{\shortstack{DeepSeek\\-Coder-6.7B}}
& Original   & 28.8 & 30.2 & 31.1 & 26.4 \\
& +LoRA      & 30.7 & 32.2 & 32.0 & 25.8 \\
& CoSec+     & 32.7 & 33.4 & 31.8 & 24.6 \\
& \textsc{CoGate} (Max-Prob)   & 34.1 & 35.7 & 37.2 & 32.4 \\
& \textsc{CoGate} (Entropy)    & \textbf{34.4} & \textbf{36.1}
             & \textbf{38.0} & \textbf{33.7} \\
\hline

\multirow{5}{*}{\shortstack{Qwen2.5\\-Coder-14B}}
& Original   & 39.3 & 41.0 & 42.1 & 37.7 \\
& +LoRA      & 41.4 & 43.0 & 43.4 & 36.8 \\
& CoSec+     & 43.8 & 44.2 & 42.5 & 37.5 \\
& \textsc{CoGate} (Max-Prob)   & 46.8 & 49.4 & 50.8 & 42.9 \\
& \textsc{CoGate} (Entropy)    & \textbf{50.1} & \textbf{56.8}
             & \textbf{54.8} & \textbf{47.0} \\
\hline

\end{tabular}
\vspace{-5pt}
\caption{CWEval Func-Sec@10 under varying
temperatures.}
\label{tab:temperature-results}
\end{table}

\noindent \textbf{Experimental Setup.}
In this section, we study the effects of sampling temperature on the performance of \textsc{CoGate} and other baselines. We select a range of temperatures $T \in \{0.4, 0.6, 0.8, 1.0\}$ and evaluate how the intervention utility and overall system behavior vary with temperature. For
CoSec+ and the two \textsc{CoGate} variants, we jointly set the base and expert temperatures to the same value, i.e., $T_B=T_S=T$. We keep all other settings fixed, including top-$p=0.95$, the CoSec+ acceptance threshold $a=0.3$,
and the family-specific gate thresholds selected on the
development split. We use Func-Sec@10 on CWEval as the metric since it measures whether a generated
program is simultaneously functionally correct and secure.

\noindent \textbf{Findings.}
From the results provided in Table~\ref{tab:temperature-results}, we observe the following: (1) The advantage of \textsc{CoGate} is minimal compared to CoSec+ under low temperatures (e.g., $T=0.4$). This is likely due to the expert distribution being relatively sharp as a result of the low temperature value, so low-confident expert proposals are uncommon and the confidence gate remains open. Therefore, there is limited benefit from preventing misleading guidance. (2) The difference between CoSec+ and \textsc{CoGate} becomes larger as the temperature increases. When the security expert distribution flattens with higher temperature, \textsc{CoGate} instead detects the low-confidence distribution and defers to the base model. This prevents the increased sampling diversity from injecting noise into the decoding process.

\section{Conclusion}
\label{sec:conclusion}

This work identifies and fixes a fundamental failure mode in co-decoding-based secure code generation: existing methods conflate whether a security expert prefers a token with whether the expert is trustworthy enough to guide generation. Our diagnosis shows that under low confidence, their acceptance rule rewards the expert's uncertainty as a security signal, generating noise on precisely the out-of-distribution inputs most critical to security. \textsc{CoGate} separates these signals through a confidence gate that lets the expert's confidence determine its authority to steer, while the relative ratio selects which token it prefers. Extensive evaluation across six models demonstrates that \textsc{CoGate} recovers the theoretical gains of co-decoding while eliminating its failure on unseen vulnerabilities. This validates our analysis and suggests that confidence-aware steering is essential for robust secure code generation.

\bibliography{aaai2027}

\end{document}